\documentclass[showpacs,preprintnumbers,amsmath,amssymb,aps,pre]{revtex4}

\usepackage{graphicx}
\usepackage{dcolumn}
\usepackage{bm}
\begin{document}

\preprint{LA-UR 08-03657}

\title{V-T Theory of  Self Dynamic Response in a Monatomic Liquid }

\author{Giulia De Lorenzi-Venneri}
\author{Eric D. Chisolm}
\author{Duane C. Wallace}

\affiliation{Theoretical Division, Los Alamos National Laboratory, 
Los Alamos, New Mexico 87545}

\date{\today}

\begin{abstract}

A new theoretical model for self dynamic response
is developed using Vibration-Transit (V-T) theory, and is applied to liquid 
sodium at all wavevectors $q$ from the hydrodynamic regime to the free 
particle limit. In this theory the zeroth-order Hamiltonian describes  the 
vibrational motion in a single random valley harmonically extended to infinity.
This Hamiltonian is tractable, is evaluated \emph{a priori} for monatomic liquids, and the
same Hamiltonian (the same set of eigenvalues and eigenvectors) is used for
equilibrium and nonequlibrium theory.  Here, for the self intermediate
scattering function $F^{s}(q,t)$, we find the vibrational contribution is in 
near perfect agreement with molecular dynamics (MD) through short and intermediate 
times, at all $q$. This is direct confirmation that normal mode vibrational correlations
are present in the motion of the liquid state. The primary transit effect is diffusive 
motion of the vibrational equilibrium positions, as the liquid transits rapidly 
among random valleys. This motion is modeled as a standard random walk, and the 
resulting theoretical $F^{s}(q,t)$ is in excellent agreement with MD results
at all $q$ and $t$. In the limit $q\rightarrow \infty$,
the theory automatically exhibits the correct approach to the free-particle limit. 
Also in the limit $q\rightarrow 0$, the hydrodynamic limit emerges as well.
In contrast to the benchmark theories of generalized hydrodynamics and mode 
coupling, the present theory is near \emph{a priori}, while achieving modestly better
accuracy. Therefore, in our view, it constitutes an improvement over the
traditional theories.

\end{abstract}

\maketitle

\section{Introduction}

Important advances have been made in the theory of equilibrium 
thermodynamic properties of liquids \cite{Ashcroft_1,Marchbook}. These advances are
characterized by the ability to calculate \textit{a priori} the measured properties of
real liquids, to an accuracy approaching the experimental accuracy itself
\cite{Ashcroft_2, Faber, Hafner}. The \textit{a priori} nature of the theory is based on the
key physical property of condensed matter, that the potential which governs
the nuclear motion is given by electronic structure theory, in the form of
the electronic groundstate energy as a function of the nuclear positions. In
applying these broad theoretical foundations, two developments have been
extremely helpful: (a) the development of pseudopotential perturbation theory
which gives effective internuclear potentials for nearly-free-electron metals
in crystal and liquid phases \cite{Harrison}, and (b) the
development of molecular dynamics (MD) computations to the point of providing
highly accurate results, indeed capable of substituting for experimental data,
when a good internuclear potential is used \cite{ARPRL74}. Based on these
principles, \textit{a priori} calculations have been made of binding energies of the
elements \cite{Ashcroft_3}, thermodynamic properties of crystals \cite{Wallace_1}
and liquids \cite{SSWPRB83}, and liquid static structure factors \cite{JKT76,ValB47}.
What is important for the present work, MD has proven its reliability for
nonequilibrium properties, e.g. $S(q,\omega)$ for crystals \cite{GHKPRB16} and liquids
\cite{ARPRL74,ValB47,JacMcDonLu}, liquid resistivity \cite{ALPR66}, 
and shear viscosity \cite{BBJCP84,CMJPF87}.

V-T theory was introduced when it was found that a theory based on two 
components of the atomic motion can account for the equilibrium thermodynamic
data of elemental liquids \cite{VT5}: (a) normal-mode vibrations in one (any)
random valley, providing $\gtrsim 90 \%$ of the thermal energy and entropy, and (b)
transits among a very large number of macroscopically equivalent random
valleys, providing the remaining $10\%$. This theory is useful because the dominant
vibrational motion is tractable,  and its Hamiltonian  can be evaluated for real
liquids from electronic structure theory \cite{VT15}. To test this theory beyond
equilibrium properties, it was applied to dynamic response in liquid sodium, where the 
vibrational contribution alone was found to give an excellent account
of experimental results for the Brillouin peak dispersion curve \cite{ARXIV05}.
A small correction for transits then produced agreement with MD data for the entire
$S(q,\omega)$ graphs \cite{GDWJCP06}. At this point we could see the possibility of a liquid
theory for both equilibrium and nonequilibrium properties based on the same
dominant (vibrational) component of the motion. Such a theory could prove
useful because traditional nonequilibrium theories primarily describe the decay
of fluctuations, through processes encoded in e.g. friction coefficients and
memory functions, and these are concepts not present in equilibrium theories.
Our point is not that the traditional description is
wrong, but that a good part of it is already contained in the same
vibrational motion that underlies the equilibrium theory. This view motivates
the present application of V-T theory to self dynamic response. This application
will provide a serious test of our theory, since self dynamic response has been 
thoroughly analyzed by the traditional theories of generalized hydrodynamics
\cite{L&VIII} and mode coupling \cite{W&S82}. These analyses
are discussed in liquid theory monographs, and have become a benchmark in dynamic response 
theories \cite{HMCDbook, BZbook}.

The previous application of V-T theory to dynamic response was evaluated
in the one-mode scattering approximation \cite{GDWJCP06}. Here, to work at larger
$q$, it is necessary to use the full vibrational theory, including normal-mode
scattering in all orders, i.e. keeping the displacement autocorrelation
functions in the exponent. This formulation has not been studied previously,
and what it reveals is remarkable to say the least. The vibrational
contribution alone, in \textit{a priori} numerical evaluation at all $q$, is in near
perfect agreement with MD calculations through short and intermediate times, and the vibrational 
contribution alone also converges to the correct theoretical
free-particle behavior at large $q$. Then, accounting for transit motion in
leading approximation produces a theory in excellent agreement with MD calculations 
at all $q$ and $t$.

In Sec.~IIA, the vibrational contribution to the self intermediate scattering
function is derived, and its short-time behavior is examined, as well as its
convergence to the free-particle limit. In Sec.~IIB, the complete time
dependence of the vibrational contribution is analyzed, and in Sec.~IIC this
contribution is compared with MD data. The transit induced correction to the 
vibrational contribution is derived and modeled in Sec.~IID. 
The hydrodynamic limit is derived in Sec. IIE.
The complete
theoretical results, vibrational plus transit, are compared with MD in Sec.~IIIA
and the two-step process by which theory approaches the free-particle limit is analyzed
in Sec.~IIIB. The salient theoretical features are summarized and discussed in
Secs.~IVA and IVB, and V-T theory is compared and contrasted with the classic benchmark
theories in Sec.~IVC.
A brief sketch of the operational procedure of V-T theory may be found in the Appendix.

\section{Theory}

\subsection{The Vibrational Contribution}

We consider a system of $N$ atoms in a cubic box at  the density of the
liquid, with periodic boundary conditions applied to the atomic motion.
The atoms are labeled $K=1,\dots,N$ and their positions are $\bm{r}_{K}(t)$ as functions
of time $t$. 
The self component of the intermediate scattering function is \cite{HMCDbook,BZbook}
\begin{equation} \label{eq1}
F^{s}(q,t) = \frac {1}{N}   \left < \sum_{K} e^{-i \bm{q}\cdot (\bm{r}_{K}(t)-\bm{r}_{K}(0))}                                        
            \right >,
\end{equation}
where the brackets indicate a motional  average in an equilibrium state.
The vibrational contribution expresses the motion in a single harmonic 
random valley extended to infinity \cite{VT5}. In this motion each atom
moves with displacement $\bm{u}_{K}(t)$ away from the fixed equilibrium
position $\bm{R}_{K}$, so that
\begin{equation} \label{eq2}
\bm{r}_{K}(t)=\bm{R}_{K} + \bm{u}_{K}(t). 
\end{equation}
Then the vibrational contribution to Eq.~(\ref{eq1}) becomes
\begin{equation} \label{eq3} 
F_{vib}^{s}(q,t)=\frac {1}{N}   \left < \sum_{K} e^{-i \bm{q}\cdot (\bm{u}_{K}(t)-\bm{u}_{K}(0))}                                       
            \right >_{vib},
\end{equation}
where $\left < \dots \right >_{vib}$ indicates an average over the vibrational motion in one
(any) random valley. This is simplified by Bloch's theorem to
\begin{equation} \label{eq4} 
F_{vib}^{s}(q,t)=\frac {1}{N}  \sum_{K}  e^{-2W_{K}(\bm{q})}\;
     e^{\left < \bm{q}\cdot\bm{u}_{K}(t) \;\;\bm{q}\cdot\bm{u}_{K}(0)\right >_{vib}}.
\end{equation}
The displacement  autocorrelation  functions are expressed in terms of the normal vibrational modes
$\lambda$ \cite {CDW07}
\begin{equation} \label{eq5} 
\left < \bm{q}\cdot\bm{u}_{K}(t)\;\; \bm{q}\cdot\bm{u}_{K}(0)\right >_{vib} =
\frac{kT}{M} \sum_{\lambda} (\bm{q}\cdot \bm{w}_{K\lambda})^{2}
\; \frac{\cos \omega_{\lambda}t}{\omega_{\lambda}^{2}},
\end{equation}
where $T$ is the temperature, $M$ is the atomic mass, $\bm{w}_{K\lambda}$ is the Cartesian vector
of the $K$ component of eigenvector $\lambda$, and $\omega_{\lambda}$ is the corresponding frequency.
The Debye-Waller factors are defined by
\begin{equation} \label{eq6} 
W_{K}(\bm{q})=\frac{1}{2} \left < (\bm{q}\cdot\bm{u}_{K}(0))^{2}\right > _{vib},
\end{equation}
and are given by Eq.~(\ref{eq5}) evaluated at $t=0$. In addition to the vibrational
averages, the right sides of Eqs.~(\ref{eq3}) and (\ref{eq4}) are to be averaged over the
allowed $\bm{q}$-vectors at each $q$-magnitude, making $F_{vib}(q,t)$ a function only of $q$, as indicated.

Important properties of $F_{vib}^{s}(q,t)$ are determined by its short-time expansion.
To find this behavior we write, from Eqs.~({\ref{eq4})-(\ref{eq6}),
\begin{equation} \label{eq7} 
\left < \bm{q}\cdot\bm{u}_{K}(t) \;\;\bm{q}\cdot\bm{u}_{K}(0)\right >_{vib} -2W_{K}(\bm{q}) = 
\frac{kT}{M} \sum_{\lambda} (\bm{q}\cdot \bm{w}_{K\lambda})^{2}
 \frac{1}{\omega_{\lambda}^{2}} (\cos \omega_{\lambda}t-1).
\end{equation}
The expansion for $\omega _{\lambda}t <<1$ is
\begin{equation} \label{eq8} 
F_{vib}^{s}(q,t) = e^{-a(q)t^{2}} \frac{1}{N}\sum_{K}e^{b_{K}(\bm{q})t^{4} - \dots}.
\end{equation}
The value at $t=0$ is
\begin{equation} \label{eq9} 
F_{vib}^{s}(q,0)=1,
\end{equation}
which is the exact theoretical result, as can be seen from Eq.~(\ref{eq1}). The
coefficient of $t^{2}$ is
\begin{equation} \label{eq10} 
a(q) = kTq^{2}/2M,
\end{equation}
obtained with the aid of the eigenvector completeness relation \cite{VT15}
\begin{equation} \label{eq11} 
\sum_{\lambda}w_{Ki,\lambda}w_{Lj,\lambda} = \delta_{KL} \delta_{ij},
\end{equation}
where $i,j$ are the Cartesian directions. In studying the time dependence, it is
advantageous to keep it in the exponent, as in Eq.~(\ref{eq8}), rather than to expand the exponential.
The leading factor in Eq.~(\ref{eq8}) is the free-particle result, defined by
\begin{equation} \label{eq12} 
F_{free}^{s}(q,t) = e^{-a(q)t^{2}}.
\end{equation}
Finally, the coefficients of $t^{4}$ in Eq.~(\ref{eq8}) are
\begin{equation} \label{eq13} 
b_{K}(\bm{q}) = \frac{kTq^{2}}{24M} \sum_{\lambda} (\hat{\bm{q}}\cdot \bm{w}_{K\lambda})^{2} \omega_{\lambda}^{2},
\end{equation}
where the unit vector $\hat{\bm{q}}$ has been introduced so as to factor out $q^{2}$.
Without approximation, the $K$-dependence of the $b_{K}(\bm{q})$ cannot be ignored.

In dynamic response, free-particle motion becomes dominant at short
times  and distances.  Free-particle behavior is contained in the self dynamic
response  \cite{HMCDbook,BZbook}, and should emerge in the large-$q$ limit of Eq.~(\ref{eq4}). Since
the right side of Eq.~(\ref{eq7}) is negative and proportional to $q^{2}$, as $q$ increases
$F_{vib}^{s}(q,t)$ will drop off more rapidly with increasing $t$.  Therefore at high $q$ only
the lowest-order term in Eq.~(\ref{eq7}) will be relevant, so that
\begin{equation} \label{eq14a} 
\lim_{q\rightarrow \infty} F_{vib}^{s}(q,t) = F_{free}^{s}(q,t).
\end{equation}
The process of the approach to this limit will be examined below.

\subsection{Free, Intermediate, and Convergence Periods}

We calculated $F_{vib}^{s}(q,t)$ for the $17\;q$ values listed in Table I. Representative 
graphs are shown in Fig.~1. Each curve has the same characteristic
shape: it starts at 1, decreases on a uniform ($q$-independent) timescale,
then levels off and converges to its long-time limit. 
The displacement autocorrelation functions, Eq.~(\ref{eq5}), decay to zero as time increases.
This decay is called the ``natural'' decorrelation,
and is responsible for the entire time dependence of $F_{vib}^{s}(q,t)$. 
Analysis reveals three 
distinct periods in the curves.

\begin{figure}[h]
\includegraphics[height=5.5in,width=3.0in]{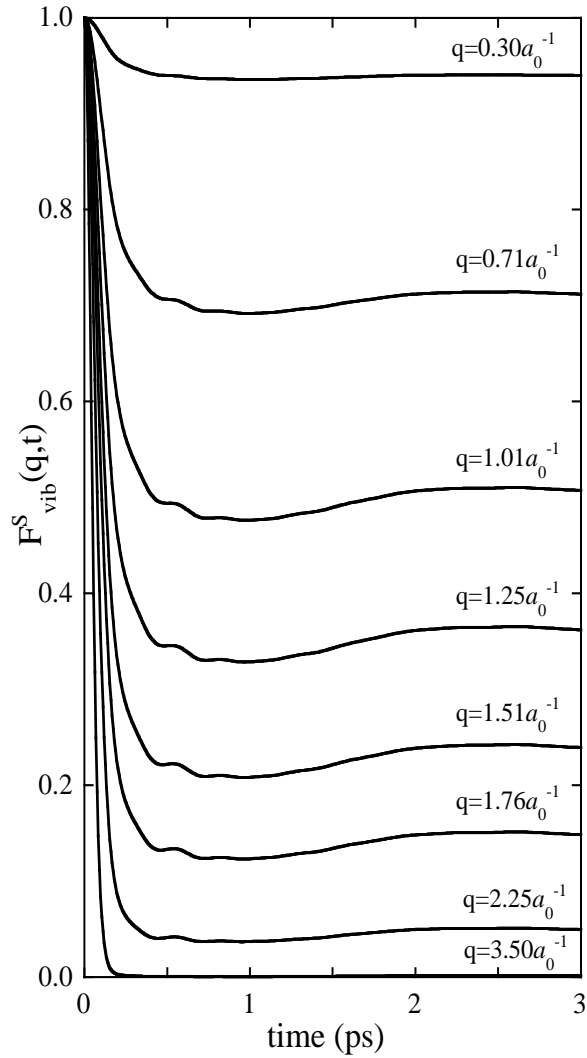}
\caption {\label{Fig1} $F^{s}_{vib}(q,t)$ for a wide range of $q$, showing the initial decrease on a
$q$-independent timescale, and universal behavior where the function converges to its
$t=\infty$ limit.}
\end{figure}

In the initial period, the atomic motion is \emph{free}. During this period,
$F_{vib}^{s}(q,t)$ is dominated by the leading factor in Eq.~(\ref{eq8}), $e^{-a(q)t^{2}}$. To
estimate the duration of this period, let us average the right side of Eq.~(\ref{eq13})
over the atoms, and over the star of $\bm{q}$; then with the eigenvector
orthonormality relation
\begin{equation} \label{eq14b} 
\sum_{Ki} w_{Ki,\lambda} w_{Ki,\lambda'} = \delta_{\lambda \lambda'}
\end{equation}
we find
\begin{equation} \label{eq15} 
\frac{1}{N}  \sum_{K} 
\left < (\hat{\bm{q}}\cdot\bm{w}_{K\lambda})^{2} \right >_{\bm{q}^{\*}} =
\frac{1}{3N}.
\end{equation}
Accordingly
\begin{equation} \label{eq16} 
F_{vib}^{s}(q,t) \approx e^{-a(q)t^{2} + b(q)t^{4} - \dots },
\end{equation}
where
\begin{equation} \label{eq17} 
b(q)=\frac{kTq^{2}}{24M} \left < \omega_{\lambda}^{2} \right >,
\end{equation}
and where $\left < \omega_{\lambda}^{2} \right > = (3N)^{-1} \sum_{\lambda} \omega_{\lambda}^{2}$. 
The free motion period will last until
the $t^{4}$ term in Eq.~(\ref{eq16}) begins to be felt; let us therefore choose $\tau_{f}$, the
duration of this period, as the time when $b(q)t^{4}=0.1 a(q)t^{2}$, giving 
$\tau_{f}=\sqrt {1.2/\left < \omega_{\lambda}^{2} \right >}$. Because of the decoupling approximation
used in deriving Eq.~(\ref{eq16}), $\tau_{f}$ is independent of $q$. 
$\left < \omega_{\lambda}^{2} \right >$ is related to the characteristic temperature $\theta_{2}$ by
$\frac{5}{3} \left <(\hbar \omega_{\lambda})^{2} \right > = (k\theta_{2})^{2}$ \cite{VT15}.
For our liquid Na system $\theta_{2}=154.1$ K \cite{GDWPRE07}, and we find $\tau_{f}=0.070$~ps. Our calculations
show that $F_{vib}^{s}(q,t)$ begins to depart from $e^{-a(q)t^{2}}$ at a time near $\tau_{f}$,
specifically at around  0.08~ps at $q=0.30~a_{0}^{-1}$, and  decreasing to around 0.05~ps at 
$q=3.50~a_{0}^{-1}$.

The intermediate
period in Fig.~1 follows the free period. Here the strong decrease of $F_{vib}^{s}(q,t)$
continues, but the function is not approximated by the $e^{-a(q)t^{2}}$ factor in
 Eq.~(\ref{eq8}). In terms of the power series expansion of the right side of
  Eq.~(\ref{eq7}), increasingly higher orders contribute while  $F_{vib}^{s}(q,t)$ retains a
smooth $t$-dependence. This property is $q$-independent.

At some point, the set of $\cos \omega_{\lambda}t$ dephase and begin to cancel, starting
from the highest frequency and continuing to the lowest, which is the last to dephase.
Since the highest 
frequency in our system is $\omega_{max}=25.5$~ps$^{-1}$ \cite{GDWJCP06}, 
 the dephasing will begin around $2\pi/\omega_{max} = 0.25$~ps. From
 Fig.~1, this is close to the uniform ($q$-independent) time when the initial 
 decrease ends and  $F_{vib}^{s}(q,t)$ begins to converge to its long-time limit. 
 Hence we associate the convergence period with the normal-mode dephasing process.

 The $t\rightarrow \infty$ limit is obtained by setting to zero the displacement autocorrelation
 functions in Eq.~(\ref{eq4}):
\begin{equation} \label{eq18} 
F_{vib}^{s}(q,\infty) = \frac{1}{N} \sum_{K}e^{-2W_{K}(\bm{q})}.
\end{equation}
In approaching this limit, when the displacement autocorrelation functions
are sufficiently small, a first-order expansion of the
time dependent part is useful. With Eq.~(\ref{eq5}) this expansion reads
\begin{eqnarray} \label{eq19} 
\lefteqn{F_{vib}^{s}(q,t) = F_{vib}^{s}(q,\infty)} \nonumber \\
 & &  +\; \frac{1}{N} \sum_{K}e^{-2W_{K}(\bm{q})}
\frac{kTq^{2}}{M} \sum_{\lambda} (\hat{\bm{q}}\cdot \bm{w}_{K\lambda})^{2} 
\frac{\cos \omega_{\lambda}t}{\omega_{\lambda}^{2}}.
\end{eqnarray}
From Eq.~(\ref{eq18}), it is the Debye-Waller factors which set the level to which
$F_{vib}^{s}(q,t)$ decreases in Fig.~1, and after that decrease,  Eq.~(\ref{eq19}) applies.
Values of $F_{vib}^{s}(q,\infty)$ are listed in Table~I.

\begin{table}
\caption{\label{table1}Infinite time limit of $F_{vib}^{s}(q,t)$ and decay factors $\gamma(q)$
of the transit induced decorrelation, according to Eq.~(\ref{eq26}), for different values of the wave vector
$q$.}
\begin{ruledtabular}
\begin{tabular}{crccc}
$q~(a_{0}^{-1})$ & $F_{vib}^{s}(q,\infty)$ &  $\gamma(q)$\\
\hline
.29711  & 0.94185  & 0.1733\\
.70726  & 0.71699  & 0.9222\\
.91575  & 0.57671  & 1.4676\\
1.0148  & 0.51138  & 1.7500\\
1.0917  & 0.46182  & 1.9753\\
1.1050  & 0.45332  & 2.0146\\
1.1443  & 0.42926  & 2.1312\\
1.2547  & 0.36482  & 2.4595\\
1.5052  & 0.24047  & 3.1805\\
1.7577  & 0.14876  & 3.8169\\
2.0041  & 0.08776  & 4.2975\\
2.2529  & 0.04903  & 4.6103\\
2.5064  & 0.02573  & 4.7423\\
2.8498  & 0.00985  & 4.6527\\
3.2000  & 0.00339  & 4.3396\\
3.5008  & 0.00126  & 3.9994\\
6.0013  & 0.00000  & 4.2271\\
\end {tabular}
\end{ruledtabular}
\end{table}

Fig.~1 shows a remarkable similarity of the curves in the convergence
period. The main feature is a broad minimum around 1~ps, where $F_{vib}^{s}(q,t)$
lies below $F_{vib}^{s}(q,\infty)$, and a final increase of $F_{vib}^{s}(q,t)$ to arrive at
$F_{vib}^{s}(q,\infty)$ by around 2~ps. These timings are accurately independent of $q$.
Furthermore, superimposed on this broad shape is a set of small oscillatory
features whose timings are also accurately independent of $q$. This
property can be understood with the aid of a small approximation in Eq.~(\ref{eq19}).
In a single random valley, the atomic sites are all inequivalent, for the same
reason that different sites in a crystal unit cell are inequivalent. If we
neglect the \emph{coupling} of this inequivalence between the eigenvectors and
Debye-Waller factors, we can average the $(\hat{\bm{q}}\cdot \bm{w}_{K\lambda})^{2}$ as in 
Eq.~(\ref{eq15}) and transform Eq.~(\ref{eq19}) to 
\begin{equation} \label{eq20} 
F_{vib}^{s}(q,t) \approx F_{vib}^{s}(q,\infty)
\;\left[1 + 
\frac{kTq^{2}}{M} \frac{1}{3N} \sum_{\lambda} 
\frac{\cos \omega_{\lambda}t}{\omega_{\lambda}^{2}} \right ].
\end{equation}
The function in brackets now exhibits uncoupled dependence on the variables
$kTq^{2}/M$ and $t$, where the $t$-dependence is parameterized by the set $\{\omega_{\lambda}\}$
of normal mode frequencies. Numerical tests verify that Eq.~(\ref{eq20}) is
rather accurate in the convergence period, and this explains the uniform
($q$-independent) time dependence  of the curves in Fig.~1.

The properties of $F_{vib}^{s}(q,t)$ were also examined for several different 
random valleys in our system, and only insignificant differences were found.
Hence the uniformity of random valleys previously found for equilibrium
thermodynamic functions \cite{GDWPRE07} is extended to the time correlation function
studied here.

\begin{figure}[t]
\includegraphics[height=5.5in,width=3.0in]{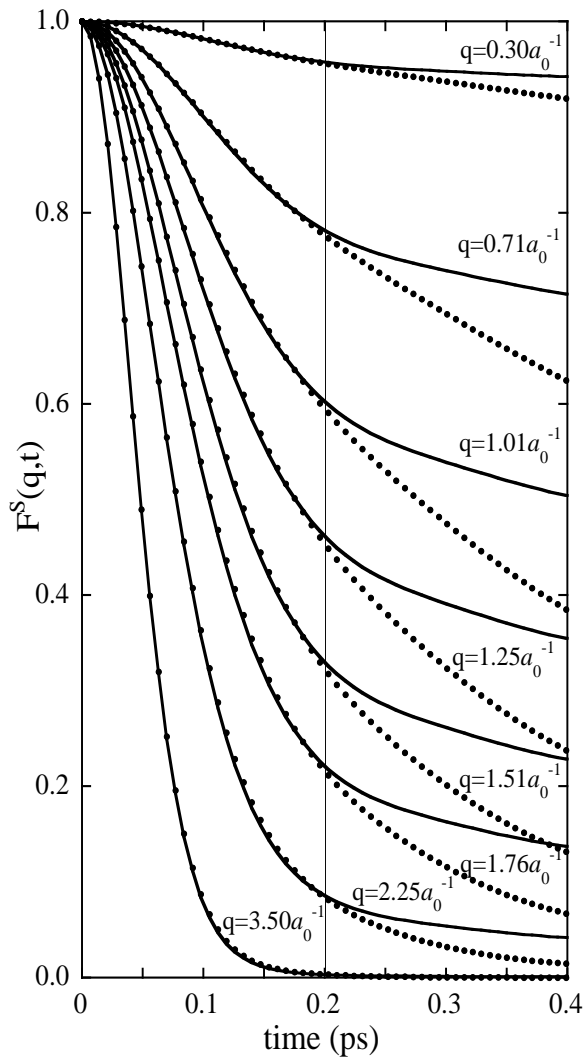}
\caption {\label{Fig2} For $F^{s}(q,t)$ the vibrational contribution (lines) and the MD data
(dots) agree extremely well to around 0.2~ps (fine vertical line).}
\end{figure}

\subsection{First Comparison with MD}

Fig.~2 shows superimposed graphs of our MD results, $F_{MD}^{s}(q,t)$, and the 
vibrational contribution, $F_{vib}^{s}(q,t)$. At each $q$, the MD and vibrational graphs
are in near perfect agreement up to an ``initial departure'' at around 0.2~ps.
Beyond the initial departure, $F_{vib}^{s}(q,t)$ turns away from $F_{MD}^{s}(q,t)$ and
proceeds to converge to its long time limit. Fig.~2 shows a remarkable
constancy of the initial departure for all $q$, and even a remarkable constancy
in the shapes of the curves in the vicinity of the departure.

Let us consider what Fig.~2 means for our analysis.  $F^{s}_{vib}(q,t)$ is based
on pure vibrational motion in a single random valley. On the other hand,
 $F^{s}_{MD}(q,t)$ is based on the real liquid motion, which has both vibrational and
 transit contributions. From Fig.~2, the vibrational motion completely dominates
  $F^{s}_{MD}(q,t)$ up to the initial departure. After that, the difference between
   $F^{s}_{MD}(q,t)$  and $F^{s}_{vib}(q,t)$ is due to transits, which are present  in the MD
   data but not in the vibrational calculation. However, even though transits are going on
   continuosly  in the MD system, at a very high rate because the temperature
   is around $T_{m}$, the effect of transits does not appear immediately in
 $F^{s}_{MD}(q,t)$, but only after the motion departs from the pure vibrational
 motion. This observation will be used to calibrate our decorrelation model
 in the next Section.

\subsection{Transit-Induced Decorrelation}

Our goal here is to model the effects of
transits in a way that is consistent with the \emph{a priori} determined vibrational
motion in a single random valley. We start by returning to the definition, Eq.~(\ref{eq1}),
which becomes
\begin{equation} \label{eq21}
F_{VT}^{s}(q,t) = \frac {1}{N}   \left < \left < \sum_{K} e^{-i \bm{q}\cdot (\bm{r}_{K}(t)-\bm{r}_{K}(0))}                                        
            \right >_{vib} \right>_{trans}.
\end{equation}
This form expresses the fundamental insight of V-T theory that the motion of the atoms consists of 
vibrations that are periodically modified by transits, allowing us to consider the effects of the two
kinds of motion as two consecutive averages.  Expanding each $\bm{r}_K$ and evaluating the vibrational
average as in Eqs.~(\ref{eq2})-(\ref{eq4}), we find 
\begin{equation} \label{eq21.5}
F_{VT}^{s}(q,t) = \frac {1}{N}  
\sum_{K}  \left< e^{-i \bm{q}\cdot (\bm{R}_{K}(t)-\bm{R}_{K}(0))} 
\; e^{-2W_{K}(\bm{q})}
     e^{\left < \bm{q}\cdot\bm{u}_{K}(t) \;\bm{q}\cdot\bm{u}_{K}(0)\right >_{vib}} \right>_{trans}.
\end{equation}
Notice that the equilibrium positions still carry time dependence because we have not yet evaluated
the transit average. From this point of view we identify two ways in which transits will modify 
$F^{s}(q,t)$. The first arises from transit-induced changes in the atomic equilibrium positions 
$\bm{R}_K(t)$, while the second way is through their effect on the displacement autocorrelation 
functions $\left < \bm{q}\cdot\bm{u}_{K}(t) \;\bm{q}\cdot\bm{u}_{K}(0)\right >_{vib}$.  Given the fact 
that transits change the equilibrium positions on very short time scales compared to the vibrational
motion, we expect the first effect to be largely decoupled from the second, so we make the approximation
of full decoupling and find
\begin{equation} \label{eq21.75}
F_{VT}^{s}(q,t) = \frac {1}{N}  
\sum_{K}  \left< e^{-i \bm{q}\cdot (\bm{R}_{K}(t)-\bm{R}_{K}(0))} \right>_{trans} \; 
\left< e^{-2W_{K}(\bm{q})}
     e^{\left < \bm{q}\cdot\bm{u}_{K}(t) \;\bm{q}\cdot\bm{u}_{K}(0)\right >_{vib}} \right>_{trans}.
\end{equation}
Now we can consider each transit average separately.

Let us abbreviate the first transit average as $A_{K}(t)$.
In V-T theory the motion of $\bm{R}_{K}(t)$ is entirely responsible for self diffusion,
hence it is appropriately modeled as a random walk.
The single atom transit rate $\nu$ is the number of transits per unit time in
which a given atom is involved. Consider an increment $\delta t$ sufficiently small
that an atom is very unlikely to be involved in more than one transit.
Then in $\delta t$, $A_K(t)$ changes by
\begin{equation} \label{eq23}
\delta A_{K}(t )= \left< \left[e^{-i \bm{q}\cdot \bm{R}_{K}(t+\delta t)}-e^{-i\bm{q}\cdot \bm{R}_{K}(t)}\right]
e^{i \bm{q}\cdot\bm{R}_{K}(0)}  \right>_{trans}.
\end{equation}
In $\delta t$, each atom transits once with probability $\nu \delta t$, or else does not
transit. If atom $K$ does transit, $\bm{R}_{K}(t+\delta t)= \bm{R}_{K}(t) +\delta \bm{R}_{K}$.
If atom $K$ does not transit, $\bm{R}_{K}(t+\delta t)= \bm{R}_{K}(t)$.  Eq.~(\ref{eq23})
becomes
\begin{equation} \label{eq24}
\delta A_{K}(t)=\left< [ e^{-i \bm{q}\cdot \delta\bm{R}_{K}}-1]
\;\;e^{-i\bm{q}\cdot (\bm{R}_{K}(t)-\bm{R}_K(0))} \right>_{trans} \nu \delta t.
\end{equation}
We assume $|\delta\bm{R}_{K}| = \delta R$, the same for all transits, while the direction of
$\delta\bm{R}_{K}$ is uniformly distributed and uncorrelated with the other factors inside
the sum. Then $[ e^{-i \bm{q}\cdot \delta\bm{R}_{K}}-1]$ can be separately averaged over angles and 
Eq.~(\ref{eq24}) can be written 
\begin{equation} \label{eq25}
\frac{\delta A_{K}(t)}{\delta t} = -\gamma (q) \left< e^{-i \bm{q}\cdot (\bm{R}_{K}(t)-\bm{R}_K(0))} 
                                       \right>_{trans}
\end{equation}
where
\begin{equation} \label{eq26}
\gamma (q) = \nu \left[ 1-\frac{\sin q\delta R}{q\delta R}\right] . 
\end{equation}
Since the transit average on the right hand side of Eq.~(\ref{eq25}) is $A_K(t)$ and $A_K(0)=1$, 
the equation integrates to
\begin{equation} \label{eq26a}
A_{K}(t)= e^{-\gamma (q) t}.
\end{equation}
Finally, from this result and Eq.~(\ref{eq4}) for $F_{vib}^{s}(q,t)$, Eq.~(\ref{eq21.75}) becomes
\begin{equation} \label{eq27}
F_{VT}^{s}(q,t) =  e^{-\gamma(q) t} \left< F_{vib}^{s}(q,t) \right>_{trans}.
\end{equation}

So the random walk for $\bm{R}_{K}(t)$ in Eq.~(\ref{eq21.75}) yields the last relation,
with the displacement autocorrelation functions in $F_{vib}^{s}(q,t)$ still having arbitrary 
time dependence.

\begin{figure}[h]
\includegraphics[height=7.0in,width=3.0in]{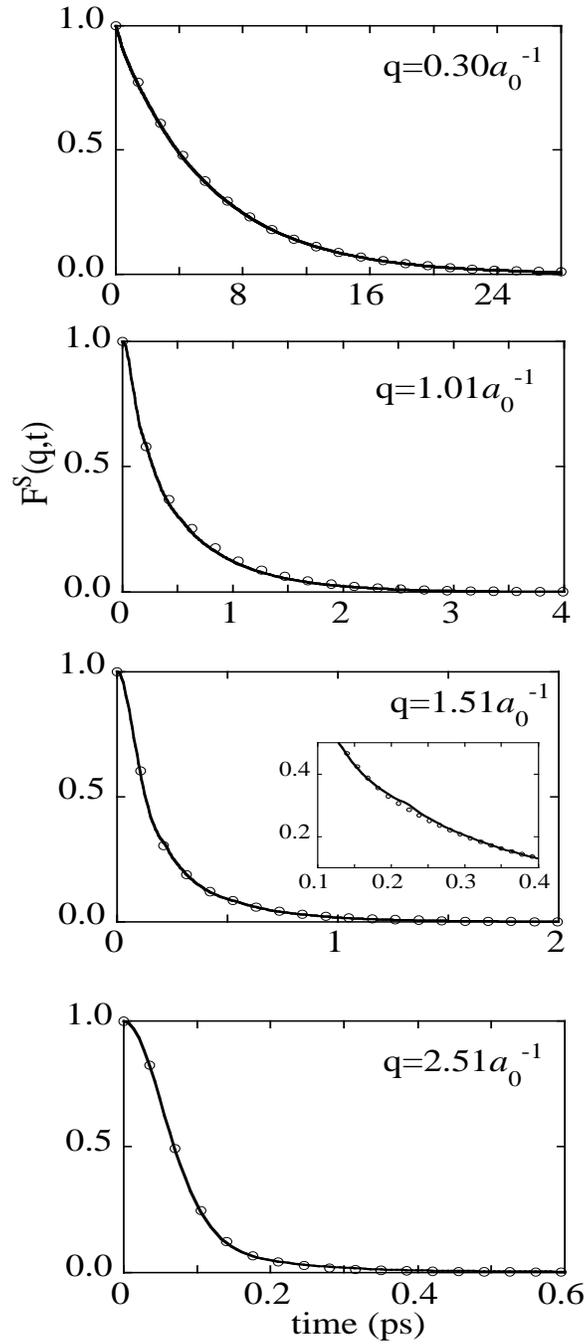}
\caption {\label{Fig3} Comparison of  $F_{VT}^{s}(q,t)$ (lines) with
$F_{MD}^{s}(q,t)$ (circles) for various $q$. The crossover bump near $\tau_{c}$ is enlarged 
in the graph for $q=1.51~a_{0}^{-1}$.}
\end{figure}

Now we must account for the fact that the transit-induced decorrelation 
does not begin to operate until some time has passed (Fig.~2). We do this
with the simplest possible model, replacing the decay factor $e^{-\gamma(q)t}$ by
the decorrelation function $D(q,t)$, defined by 
\begin{equation} \label{eq28}
 D(q,t) = \begin{cases}
  1                   &\text{for $t\leq \tau_{c}$,} \\
 e^{-\gamma(q)(t-\tau_{c})} &\text{for $t\geq \tau_{c}$;} 
\end{cases}
\end{equation}
where $\tau_{c}$ is the crossover time, to be calibrated. Then from Eq.~(\ref{eq27}), 
our transit decorrelation model is now
\begin{equation} \label{eq29}
F_{VT}^{s}(q,t)=\left< F_{vib}^{s}(q,t) \right>_{trans} D(q,t).
\end{equation}

In the random walk model,
the self diffusion coefficient is $D= \frac{1}{6} \nu (\delta R)^{2}$. A measurement of $\delta R$ 
was obtained  from MD simulations at 30~K \cite{VT11b}, and the result can be used here because 
$\delta R$ has only weak $T$ dependence. On the other hand, $\nu$ is a strong function of $T$,
so we use the MD evaluation $D_{MD} = 5.61 (10^{-5}\text{cm}^{2}/\text{s})$ at the temperature
of the present study, 395~K \cite{VT8}. For comparison,
the experimental value for  liquid Na at 395~K is 
$ 5.08 (10^{-5}\text{cm}^{2}/\text{s})$ \cite{Naexp}. The results for $\delta R$ and $\nu$ are
\begin{eqnarray} 
&\delta R &=1.75 ~a_{0},\nonumber \\
&\nu      &=3.9 \text{~ps}^{-1}. \nonumber
\end{eqnarray}
The value of $\nu$ is not far from our previous estimate of 2.5~ps$^{-1}$ \cite{DWPRE01}.
Data for $\gamma(q)$ are listed in Table~I.

The second effect of transits is felt on the displacement correlation functions.  As
 noted in Sec.~IIB, these functions already decay to zero by the natural decorrelation
 among normal modes.  As a model, we might suppose that
 transits cause sudden vibrational phase shifts, and therefore enhance the
 decay of the displacement autocorrelation functions. This enhanced decay will cause
  $F_{vib}^{s}(q,t)$  to decay more quickly to  $F_{vib}^{s}(q,\infty)$. Referring to the curves in
  Fig.~1, the main effect will be to slightly raise the broad minimum around 1~ps, and bring the curve to
    $F_{vib}^{s}(q,\infty)$  noticeably before 2~ps. 
	This is an interesting physical effect, but is small; in fact it is
	extremely small compared to the decorrelation caused by the motion of the
	equilibrium positions, as modeled in Eq.~(\ref{eq28}). We therefore neglect the
	transit-induced decorrelation of the displacement autocorrelation functions in the
	present study, with the final result that
\begin{equation} \label{eq29.5}
F_{VT}^{s}(q,t)=F_{vib}^{s}(q,t) D(q,t).
\end{equation}

Now the only thing that remains to be determined is the parameter $\tau_{c}$.
The foremost property revealed in Fig.~2 is that $\tau_{c}$ must be independent 
of $q$. For this reason it is conceivable that vibrational information alone
can determine $\tau_{c}$. We have clues to this effect -- for example, $\tau_{c}$ is
close to the start of dephasing at 0.25~ps (Sec.IIB) -- but we have no
certainty. We therefore calibrate $\tau_{c}$ by comparison of the theoretical 
$F_{VT}^{s}(q,t)$, Eq.~(\ref{eq29.5}), with the MD results in Fig.~2, at $t$ around $\tau_{c}$.
This gives $\tau_{c}=0.22$~ps. The reason $\tau_{c}$ is a little larger than the initial departure time 
of 0.20~ps in Fig.~2 is to allow the discontinuous $D(q,t)$, Eq.~(\ref{eq28}),
 to better fit the actual smooth crossover behavior. When $\tau_{c}$  is
 determined in the same way for different random valleys, the scatter in $\tau_{c}$
 is around $\pm 0.01$~ps.

\begin{figure}[t]
\includegraphics[height=2.5in,width=3.0in]{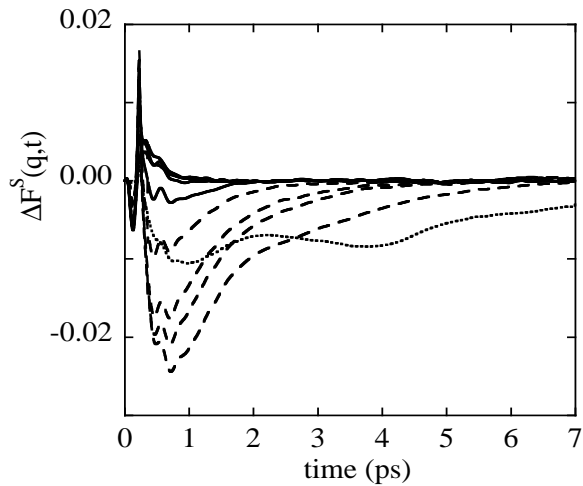}
\caption {\label{Fig4} Deviations of $F^{s}_{VT}(q,t)$ from $F^{s}_{MD}(q,t)$, 
Eq.~(\ref{eq31}). The dotted line is $q=0.30~a_{0}^{-1}$, the four dashed lines are (from the
lowest) $q=0.71, 0.92, 1.01, \text{and} 1.25 ~a_{0}^{-1}$, and the solid lines show negligible deviation
for $q\geq 1.51~a_{0}^{-1}$.}
\end{figure}

\begin{figure}[t]
\includegraphics[height=7.0in,width=3.0in]{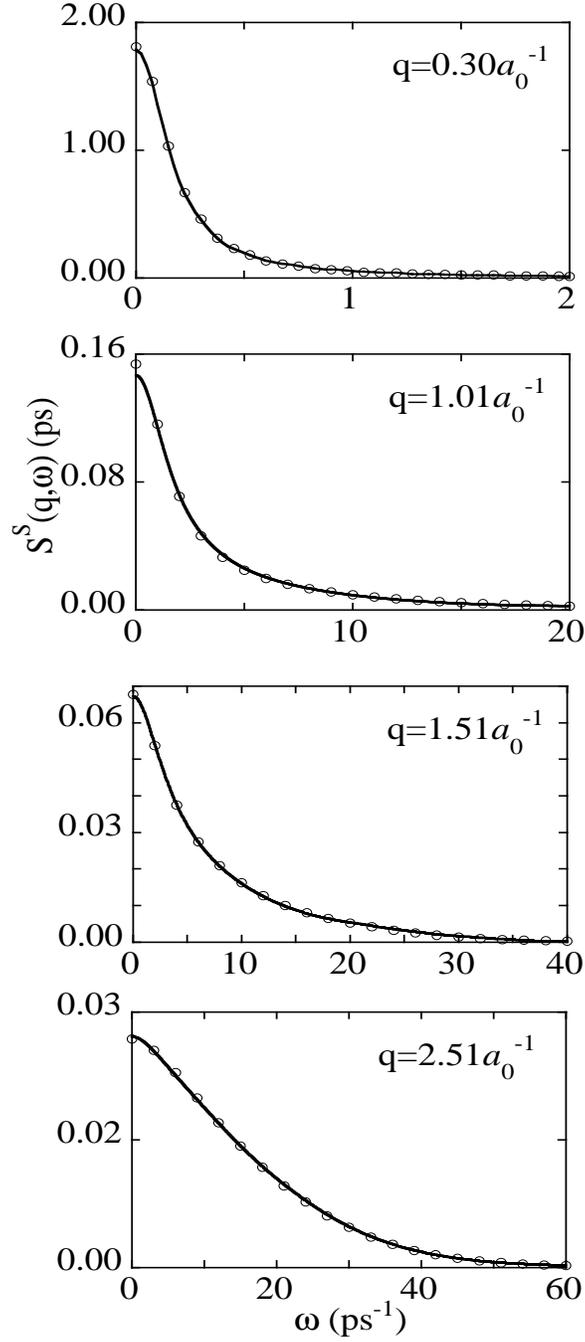}
\caption {\label{Fig5} Comparison of $S^{s}_{VT}(q,\omega)$ (lines) with
$S_{MD}^{s}(q,\omega)$ (circles) for various $q$.}
\end{figure}

\subsection{The Hydrodynamic Limit}

The hydrodynamic limit corresponds to $q \rightarrow 0$ and $t$ larger than a characteristic time.
As $q \rightarrow 0$, from Eqs.~(\ref{eq4})-(\ref{eq6}), $F_{vib}^{s}(q,t) \rightarrow 1+O(q^{2}) $ for
$0 \leq t \leq \infty$.
Also as $q \rightarrow 0$, $\gamma(q) \rightarrow O(q^{2})$, and from Eq.~(\ref{eq28}), $D(q,t) \rightarrow 
e^{-\gamma(q)t} [1+O(q^{2})]$ for $t > \tau_{c}$. Then from Eq.~(\ref{eq29.5}), $F_{VT}^{s}(q,t)$ is 
$e^{-\gamma(q) t} [1+O(q^{2})]$. We drop the
term in $O(q^{2})$ and take the  $q \rightarrow 0$ limit of $\gamma(q)$, Eq.~(\ref{eq26}), to find
\begin{equation} \label{eq30}
\lim_{q\rightarrow 0} F_{VT}^{s}(q,t) = e^{-Dq^{2}t}  \;\; \mbox{for\;\; $t > \tau_{c}$},
\end{equation}
where we used the random walk model expression for the self diffusion coefficient.
The right side is the correct hydrodynamic limit of $F^{s}(q,t)$.

\section{Comparison of Theory with MD calculations}

\subsection{Self Dynamic Response}

\begin{figure}[t]
\includegraphics[height=2.5in,width=3.0in]{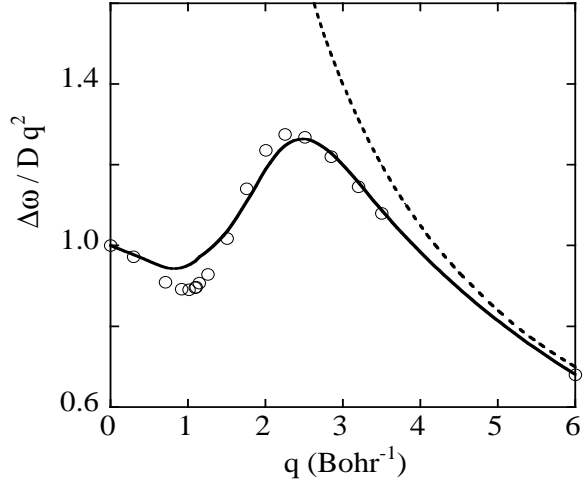}
\caption {\label{Fig6} Comparison of theory (solid line) and MD (circles) for the normalized
halfwidth of $S^{s}(q,\omega)$ for all $q$ listed in Table I. 
The dashed line is the free-particle limit.}
\end{figure}

\begin{figure}[h]
\includegraphics[height=2.5in,width=3.0in]{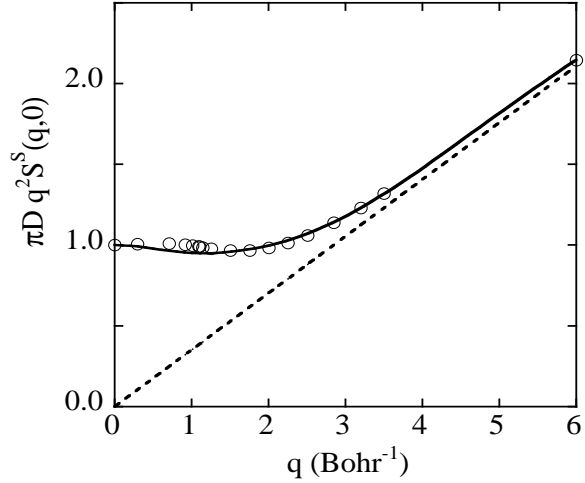}
\caption {\label{Fig7} Comparison of theory (solid line) and MD (circles) for the normalized
value of  $S^{s}(q,0)$ for all $q$ listed in Table I. 
The dashed line is the free-particle limit.}
\end{figure}

Graphs comparing theory with MD are shown in Fig.~3 for $q=0.30-2.51 ~a_{0}^{-1}$. 
The time required to decay to zero decreases strongly as $q$ increases, from around 28~ps
at $q=0.30~a_{0}^{-1}$ to 0.6~ps at  $q=2.51~a_{0}^{-1}$. The decay rates $\gamma(q)$, Table~I,
show a corresponding strong increase as $q$ increases. Since the
decorrelation begins at $\tau_{c}=0.22$~ps, nearly the entire curve at  $q=0.30~a_{0}^{-1}$
is the decay process. The decorrelation steadily becomes a smaller part of the curve as $q$
increases, until it affects only the tail of the curve at $q=2.51~a_{0}^{-1}$. These changes 
are accompanied by a change in shape of the graphs, from a near-exponential at 
$q=0.30~a_{0}^{-1}$, going over to a qualitatively Gaussian shape at $q=2.51~a_{0}^{-1}$.
All of this behavior has the important consequence that the decay rate $\gamma(q)$  eventually
becomes irrelevant as $q$ increases.

Generally speaking, the crossover bump in the theoretical curve around
$\tau_{c}$ cannot be seen in graphs as small as those in Fig.~3. 
The prominence of the bump depends on both the slope and magnitude
of the curve at $\tau_{c}$.
It is most 
prominent at $q=1.51~a_{0}^{-1}$, where it is enlarged in the inset. The deviations
\begin{equation} \label{eq31}
\Delta F^{s} (q,t) = F_{VT}^{s}(q,t) -  F_{MD}^{s}(q,t)
\end{equation}
for representative $q$ are shown in Fig.~4.  The largest deviations in our
study are included in the figure. For most $q$ and most $t$, the deviation
is $\lesssim 0.005$ in magnitude, which we consider insignificant. Larger
deviations are seen in two places.
(a) The positive spike at $t\approx 0.22$~ps is due to the crossover model, Eq.~(\ref{eq28}). This error 
could be removed by smoothing, but since it is so
small we are willing to forego the introduction of a smoothing model.
(b) The only significant deviations are those in the negative dip at
$t\sim 0.3-2.0$~ps. These deviations reach magnitude $\sim 0.01-0.02$ for the
dotted curve  ($q=0.30~a_{0}^{-1}$) and the four dashed curves  ($q=0.71-1.14~a_{0}^{-1}$).
Notice the shape at $q=0.30~a_{0}^{-1}$ is the same as the others, but stretched out
over a much longer time. This error arises from the broad shallow
minimum in $F_{vib}^{s}(q,t)$ observed in Fig.~1. That minimum is not
entirely in error, however, since a weakened copy of it is present in
$F_{MD}^{s}(q,t)$. From a detailed examination of the data, we conclude that
the error can be eliminated by adding a theoretical model for the transit-induced
decorrelation of the displacement autocorrelation functions. This decorrelation process
was discussed at the end of Sec.~IID.

Graphs comparing theory with MD for $S^{s}(q,\omega)$ are shown in Fig.~5.
The only significant error corresponds to the negative dip in $\Delta F^{s}(q,t)$
for a limited $q$ range, in Fig.~4. This $q$ range is near the location of the first peak in
$S(q)$, at $q_{m}=1.05~a_{0}^{-1}$ in our system.

Further comparison of theory and MD is shown in Figs.~6 and 7. The
quantity graphed  in Fig.~6 is $\Delta \omega(q)/Dq^{2}$, where $\Delta \omega(q)$ is the halfwidth of $S^{s}(q,\omega)$
and $Dq^{2}$ is the halfwidth in the diffusion limit \cite{HMCDbook, BZbook}.
Fig.~7 shows the quantity $\pi Dq^{2} S^{s}(q,0)$, where  $(\pi Dq^{2})^{-1}$ is
$S^{s}(q,0)$ in the diffusion limit \cite{L&VIII}. In these graphs we used
the same value $D_{MD}$ for Na at 395~K as in calibrating $\gamma(q)$ following 
Eq.~(\ref{eq29.5}). These figures will allow  a comparison of the present theory with the 
traditional generalized hydrodynamics and mode coupling theories in Sec.~IV. 
Here we observe that Figs.~6 and 7 show a rather good agreement 
between theory and MD for the entire $q$ range, including the approach to
the free-particle limit. Further, most of the error revealed in Figs.~6 and 7
appears in the vicinity of $q_{m}$ and is attributed to our neglect of 
the transit-induced decorrelation of
the displacement autocorrelation functions in Eq.~(4).

\subsection{Approach to the Free-Particle Limit}

According to Eq.~(\ref{eq14a}), $F_{vib}^{s}(q,t)$ reaches the free-particle limit as $q\rightarrow \infty$.
When the transit-induced motion of the equilibrium positions is accounted
for, the V-T theory expression is Eq.~(\ref{eq29.5}), with $D(q,t)$ given by Eq.~(\ref{eq28}).
The approach of $F_{VT}^{s}(q,t)$ to the free-particle limit is therefore to be
understood by watching the function $F_{vib}^{s}(q,t) D(q,t)$ as $q$ increases.
The process contains two steps, both operating at all times, but in
effect more or less sequential, as follows. 

(a) In the first step, because $F_{vib}^{s}(q,t)$ decreases to zero in an ever
shorter time as $q$ increases, and because $D(q,t)=1$ for $t\leq \tau_{c}$, the
decorrelation function effectively approaches $1$ as $q$ increases.  This process
is apparent in Fig.~2. Let us define $q_{c}$ by the condition that
$D(q,t)$ can be replaced by $1$ for $q \geq q_{c}$. Then
\begin{equation} \label {eq32}
F_{VT}^{s}(q,t)=F_{vib}^{s}(q,t), \;\; \mbox{for\;\;$ q\geq q_{c}$}.
\end{equation}
This is the situation in Fig.~8 at $q=3.50~a_{0}^{-1}$:
$D(q,t)$ is effectively $1$ for all time, and as a result, $F_{VT}^{s}$ and $F_{MD}^{s}$ are
in near perfect agreement for all time.

\begin{figure}[h]
\includegraphics[height=2.5in,width=3.0in]{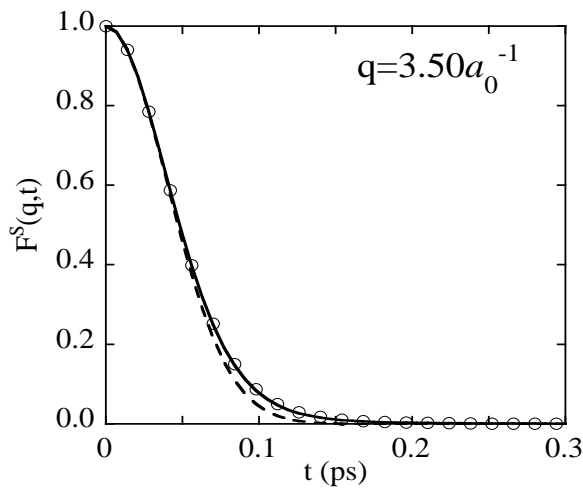}
\caption {\label{Fig8} $F^{s}(q,t)$ at $q \approx q_{c}$. Solid line is V-T theory, circles are MD, and
 dashed line is the free-particle theory, Eq.~(\ref{eq12}).}
\end{figure}
\begin{figure}[b]
\includegraphics[height=2.5in,width=3.0in]{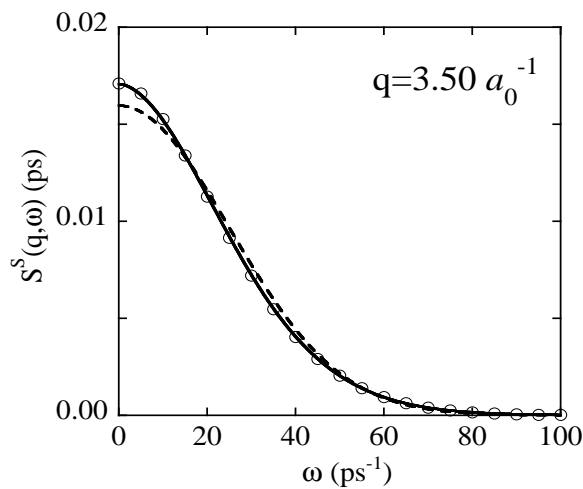}
\caption {\label{Fig9} $S^{s}(q,\omega)$ at $q \approx q_{c}$. Solid line is V-T theory, circles are MD, and
 dashed line is the free-particle theory.}
\end{figure}

(b) In exact free-particle motion, the self intermediate scattering function is
given by Eq.~(\ref{eq12}). But when step (a) is completed and Eq.~(\ref{eq32})
holds, and $F_{vib}^{s}(q,t)$ agrees with $F_{MD}^{s}(q,t)$, these functions have not yet
reached the free-particle limit. This is also shown in Fig.~8, where the
theory and MD curves differ noticeably from the free-particle curve. The
situation exemplifies the point discussed in Sec.~IIB, that in the intermediate time period,
$F_{vib}^{s}(q,t)$ continues its steep decrease but is
not well approximated by the $e^{-a(q)t^{2}}$ factor. So the final step in
arriving at the free-particle limit is to increase $q$ beyond $q_{c}$, until
$F_{vib}^{s}(q,t)$ concides with $F_{free}^{s}(q,t)$.

Fig.~9 shows the Fourier transform of Fig.~8. Again the theory and MD are
in agreement while both differ from the free-particle curve. Another view of 
the approach to the free-particle limit is seen in Fig.~6. For the halfwidth,
the difference between theory and MD is around $0.8\%$ for $q=2.51-3.50~a_{0}^{-1}$, and 
is  $0.1\%$ at $q=6.00~a_{0}^{-1}$. This level of agreement is in strong contrast to the
much larger difference of theory and MD from the free-particle halfwidth.
The same property for $S^{s}(q,0)$ is observed in Fig.~7. Clearly theory and
MD are in substantial agreement long before they arrive at the free-particle limit.
Then, the final approach to the free particle limit is determined entirely by the vibrational 
contribution, and is very slow, possibly algebraic, as seen in Figs.~6 and 7.

\section{Summary and Conclusions}

\subsection{Vibrational Contribution}

The main conclusions of this study are expressed through a discussion of the
self intermediate scattering function. We begin by comparing
properties of $F_{vib}(q,t)$ with exact theory and with MD results.

(a) The short-time expansion is Eqs.~(\ref{eq8}), (\ref{eq10}), and (\ref{eq13}).
The value at $t=0$ is $F_{vib}^{s}(q,0)=1$, the exact result. The expansion gives the free-particle
limit $F_{free}^{s}(q,t)=e^{-a(q)t^{2}}$ as the leading factor in $F_{vib}^{s}(q,t)$, at all $q$. 
The result is
not trivial. It appears because  the eigenvector completeness relation Eq.~(\ref{eq11}) decouples
the free-particle motion in the exponent. This property ensures that V-T theory will automatically
give the free-particle behavior as $q\rightarrow \infty$.

(b) At the end of the period of free-particle motion, at $t=\tau_{f}$, $F_{vib}^{s}(q,t)$
starts to depart from $e^{-a(q)t^{2}}$. $F_{vib}^{s}(q,t)$ continues its strong decrease
throughout the intermediate period, until $F_{vib}^{s}(q,t)$ levels off and
starts to converge to $F_{vib}^{s}(q,\infty)$. The power series in $t^{2}$ does not usefully
represent $F_{vib}^{s}(q,t)$ in the intermediate period. $F_{vib}^{s}(q,t)$ is in very good agreement 
with $F_{MD}^{s}(q,t)$ for all times up to near the end of the intermediate period, and all $q$, Fig.~2. 
This period of agreement between  $F_{vib}^{s}(q,t)$ and $F_{MD}^{s}(q,t)$ is three times longer than
the duration $\tau_{f}$ of free-particle motion, providing direct confirmation that
normal mode vibrational correlations are present in the motion of the liquid state.

\subsection{Transit Contribution}

To complete the theory for the self intermediate scattering function, we must 
include the transit motion. What is needed is a physically motivated model
which will not interfere with the above listed accurate properties of the
vibrational contribution. In Eq.~(\ref{eq4}) for $F_{vib}^{s}(q,t)$, there are two places where
transits will cause decorrelation of the purely vibrational atomic motion.

(a) The first effect of transits is to make the equilibrium positions 
time-dependent in Eq.~(\ref{eq2}), i.e. $\bm{R}_{K}(t)$. This motion is modeled as a random
walk of transit jumps $\delta \bm{R}_{K}$, all of the same magnitude $\delta R$, and uniformly
distributed over angles for each atom. This transit motion then decouples 
from the vibrational motion, to give the complete theory in the form of Eqs.~(\ref{eq28}) 
and (\ref{eq29.5}).  The random walk model is calibrated \textit{a priori} from MD data for the transit jump
distance and the self diffusion coefficient. The crossover time $\tau_{c}$ is the only parameter
calibrated with the aid of MD data for self intermediate scattering. The transit decorrelation embodied
in $D(q,t)$ constitutes a major correction to $F_{vib}^{s}(q,t)$, and brings $F_{VT}^{s}(q,t)$ into 
excellent overall agreement with $F_{MD}^{s}(q,t)$.

(b) The displacement autocorrelation functions, written in Eq.~(\ref{eq5}), decay
to zero as $t\rightarrow \infty$, by virtue of the natural decorrelation among normal mode
vibrations. The second transit effect will be to \emph{enhance} this decay. 
To make a proper model for this will require additional study, especially regarding the form of the decorrelation
and the importance of the $K$ dependence. In the meantime, the present study shows that this transit
effect is small enough to neglect entirely and still have a highly accurate theory.

For the self function, the transition to free-particle behavior involves two
sequential steps. First, with increasing $q$, $F_{vib}^{s}(q,t)$ goes to zero in a shorter
and shorter time, until at $q_{c}$ $F_{vib}^{s}(q,t)$ vanishes for $t \gtrsim \tau_{c}$. Then for
$q \geq q_{c}$, $D(q,t)$ in Eq.~(\ref{eq29.5}) is effectively 1, so that $F_{VT}^{s}(q,t)=F_{vib}^{s}(q,t)$,
and $F_{VT}^{s}(q,t)$ agrees with $F_{MD}^{s}(q,t)$. As $q$ increases from $q_{c}$, theory and MD
remain in agreement, and together they approach the free-particle limit. 
This final approach is within the vibrational contribution.

In contrast, the hydrodynamic limit depends entirely on the transit decorrelation. As
$q \rightarrow 0$, $F_{VT}^{s}(q,t) \rightarrow e^{-\gamma(q) t}$ for $t > \tau_{c}$, and this produces the
hydrodynamic limit in Eq.~(\ref{eq30}).

\begin{figure}[h]
\includegraphics[height=2.5in,width=3.0in]{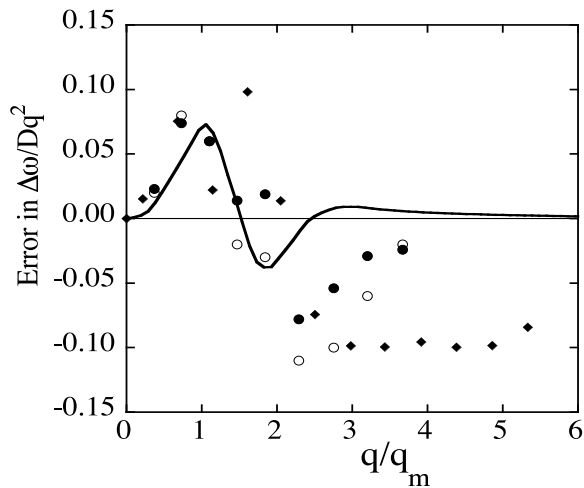}
\caption {\label{Fig10} Error in $\Delta \omega /Dq^{2}$ as function of $q/q_{m}$. Line is present theory,
($q_{m}=1.05~a_{0}^{-1}$), open and filled circles are Ar from \cite{L&VIII} and \cite{W&S82} respectively,
and filled diamonds are Rb from \cite{W&S82}.}
\end{figure}
\begin{figure}[h]
\includegraphics[height=2.5in,width=3.0in]{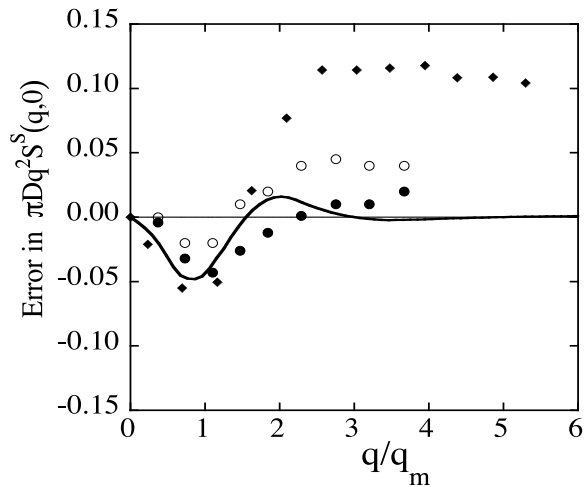}
\caption {\label{Fig11} Error in $\pi Dq^{2} S^{s}(q,0)$ as function of $q/q_{m}$. Line is present theory,
($q_{m}=1.05~a_{0}^{-1}$), open and filled circles are Ar from \cite{L&VIII} and \cite{W&S82} respectively,
and filled diamonds are Rb from \cite{W&S82}.}
\end{figure}

\subsection{Comparison with Traditional Theories}

The self dynamic structure factor for liquid Ar near the triple point was
studied by Levesque and Verlet \cite{L&VIII}, who compared several generalized
hydrodynamics models with MD for the quantities graphed in Figs.~6 and 7. 
A similar study was made for liquids Ar and Rb by Wahnstr{\"o}m and Sj{\"o}gren 
\cite{W&S82}, who carried out their analysis in mode coupling theory. These studies
are discussed by Hansen and McDonald \cite{HMCDbook} (p.~266) and by Balucani and Zoppi 
\cite{BZbook} (Sec.~5.3), and have become the benchmark theories for $S^{s}(q,\omega)$. 
Levesque and Verlet use the memory function formalism ``to give a simple
phenomenological fit for the computed self intermediate scattering function.'' 
The fitting process uses the $q$-dependent coefficients of $t^{2}, t^{4}$, and $t^{6}$ in
$F^{s}(q,t)$, and is also adjusted to give the correct ideal gas (high-$q$) limit.
Wahnstr{\"o}m and Sj{\"o}gren split the memory function into a rapid binary
collision part and a slowly varying collective recollision part. Their calibration
also uses the coefficients of $t^{2}, t^{4}$, and $t^{6}$ in $F^{s}(q,t)$, and additional
longer time information. The application of V-T theory constitutes an extreme contrast
to these traditional theories. The collision concept does not appear in
V-T theory. The vibrational contribution alone, \textit{a priori} and for all $q$, accurately
accounts for the short- and intermediate-time behavior of $F^{s}(q,t)$ and 
accurately accounts for the free-particle limit as well. Hence $F_{vib}^{s}(q,t)$ provides
much of the information used for calibration by the traditional theories. Then,
only the transit contribution remains to be addressed. 


Since we are proposing a radically new theory of self dynamic response, we
should compare numerical accuracy with the traditional theories. The standard
for this comparison is the error of theory from MD for the quantities graphed in Figs.~6 and 7,
the width and height of $S^{s}(q,\omega)$ \cite{L&VIII,W&S82,HMCDbook,BZbook}. These errors,
in the form of (theory-MD)/MD, are compared for traditional and present theories
in Figs.~10 and 11. The Levesque and Verlet
data for Ar are from their tables \cite{L&VIII}, and the  Wahnstr{\"o}m and Sj{\"o}gren 
data for Ar and Rb are read from their graphs \cite{W&S82}. The large error for Rb at large 
$q$ is attributed to a difference of the self diffusion coefficient between 
theory and MD \cite{W&S82}. In the overall comparison, the error in the present theory
is smaller than in the traditional theories. However, in our view, more
important than the comparison of numerical accuracies is the near \emph{a priori}
character of the present theory. Beyond the standard diffusional model for 
the motion of the vibrational equilibrium positions, only the single scalar parameter 
$\tau_{c}$ is needed to calibrate $F_{VT}^{s}(q,t)$ for all $q$ and $t$.


\appendix
\section{Operational Procedures of V-T theory}

Since the operational procedures of V-T theory are somewhat novel, a brief 
summary will perhaps be useful. One begins with an interatomic potential for
the system of interest. An MD system is constructed, and quenches are made from the liquid 
to minimum potential energy structures, where the equilibrium positions
and the dynamical matrix are evaluated. Except for three translational 
eigenvalues, which are zero to numerical accuracy, all normal mode eigenvalues
must be positive. In the first application to a given system, one needs to make 
some tests to identify the random structures (as differentiated from symmetric
structures), and to verify that the random structures do indeed dominate the
potential energy surface \cite{VT8,VT9,GDWPRE07}. This step is to check the ``single random
valley" approximation, whose verification is still in progress \cite{GDWPRE07}. From this
point one can proceed by working with a single random valley. The dynamical 
matrix is diagonalized to find the frequencies $\omega_{\lambda}$ and eigenvectors $\bm{w}_{K\lambda}$.
The vibrational contribution to any thermodynamic function \cite{VT5,VT15} or to a time
correlation function \cite{CDW07} can be expressed in terms of these quantities. This
is illustrated by the equations of Sec.~II. Numerical evaluation of sums 
$\sum_{\lambda}f_{\lambda}$ proceeds by direct summation over the $3N-3$ modes with nonzero 
$\omega_{\lambda}$. At the present stage in the theoretical development, the transit
contribution to a statistical mechanical average is accounted for by a
macroscopic model, such as the one reported here in Sec.~IID.

\acknowledgments{Brad Clements and Renzo Vallauri are gratefully acknowledged 
for helpful discussions.  This work was supported by the U.\ S. DOE under Contract No.\ DE-AC52-06NA25396.}

\bibliography{TheoryforFs} 

\end{document}